\documentclass[conference]{IEEEtran}
\IEEEoverridecommandlockouts
\usepackage{cite}
\usepackage{amsmath,amssymb,amsfonts}
\usepackage{algorithmic}
\usepackage{graphicx}
\usepackage{textcomp}
\usepackage{xcolor}

\usepackage{booktabs}

\usepackage{multirow}

\usepackage[super]{nth}

\usepackage{acronym}
\newacro{sme}[SME]{Small and Medium-sized Enterprise}
\newacro{it}[IT]{Information Technology}
\newacro{ot}[OT]{Operation Technology}
\newacro{cps}[CPS]{Cyber-Physical System}
\newacro{cpps}[CPPS]{Cyber-Physical Production System}
\newacro{ids}[IDS]{Intrusion Detection System}
\newacro{svm}[\textit{SVM}]{\textit{Support Vector Machine}}
\newacro{wsn}[WSN]{Wireless Sensor Network}
\newacro{darpa}[DARPA]{Defense Advanced Research Projects Agency}
\newacro{kdd}[KDD]{Knowledge Discovery in Databases}
\newacro{scada}[SCADA]{Supervisory Control And Data Acquisition}
\newacro{dpi}[DPI]{Deep Packet Inspection}
\newacro{dmz}[DMZ]{De-Militarized Zone}
\newacro{iot}[IoT]{Internet of Things}
\newacro{cc}[C\&C]{Command and Control}
\newacro{wlan}[WLAN]{Wireless Local Area Network}
\newacro{gui}[GUI]{Graphical User Interface}
\newacro{cli}[CLI]{Command Line Interface}
\newacro{wui}[WUI]{Web-based User Interface}
\newacro{api}[API]{Application Programming Interface}

\def\BibTeX{{\rm B\kern-.05em{\sc i\kern-.025em b}\kern-.08em
    T\kern-.1667em\lower.7ex\hbox{E}\kern-.125emX}}
\begin{document}

\title{Investigating the Ecosystem of Offensive Information Security Tools\\
\thanks{This work has been supported by the German Federal Ministry of Education and Research (BMBF) (Foerderkennzeichen 01IS18062E, SCRATCh). The authors alone are responsible for the content of the paper.
This work is a preprint of a paper accepted at the 1st Workshop on Next Generation Networks and Applications (NGNA-2020). Please cite as:
\textbf{\textit{SD Duque Anton, D Fraunholz, and D Schneider. ``Investigating the Ecosystem of Offensive Information Security Tools'', in: 1st Workshop on Next Generation Networks and Applications (NGNA-2020), Kaiserslautern, Germany (December 2020)}} 
}
}

\author{\IEEEauthorblockN{Simon D Duque Anton\IEEEauthorrefmark{1}\IEEEauthorrefmark{2}, Daniel Fraunholz\IEEEauthorrefmark{1}\IEEEauthorrefmark{2},
and Daniel Schneider\IEEEauthorrefmark{1}\IEEEauthorrefmark{2}
}\\
\IEEEauthorblockA{\IEEEauthorrefmark{1}\textit{Intelligent Networks Research Group} \\
\textit{German Research Center for Artificial Intelligence}\\
Kaiserslautern, Germany \\ \\}
\IEEEauthorblockA{\IEEEauthorrefmark{2}\textit{Chair for Wireless Communication and Navigation} \\
\textit{University of Kaiserslautern}\\
Kaiserslautern, Germany \\\\
Email: \{firstname\}.\{lastname\}@dfki.de}
}

\maketitle

\begin{abstract}
The internet landscape is growing and at the same time becoming more heterogeneous.
Services are performed via computers and networks,
critical data is stored digitally.
This enables freedom for the user,
and flexibility for operators.
Data is easier to manage and distribute.
However,
every device connected to a network is potentially susceptible to cyber attacks.
Security solutions,
such as antivirus software or firewalls,
are widely established.
However,
certain types of attacks cannot be prevented with defensive measures alone.
Offensive security describes the practice of security professionals using methods and tools of cyber criminals.
This allows them to find vulnerabilities before they become the point of entry in a real attack.
Furthermore,
following the methods of cyber criminals enables security professionals to adapt to a criminal's point of view and potentially discover attack angles formerly ignored.
As cyber criminals often employ freely available security tools,
having knowledge about these provides additional insight for professionals.
This work categorises and compares tools regarding metrics concerning maintainability,
usability and technical details.
Generally,
several well-established tools are available for the first phases,
while phases after the initial breach lack a variety of tools.
\end{abstract}


\section{Introduction}
The \ac{iot} provides a number of benefits,
to consumers as well as business organisations.
Flexibility and ease of use,
coupled with low management overhead and cost make it a desirable concept.
Smart cities,
for example,
can benefit from \ac{iot} solutions~\cite{formisano2015advantages},
as well as cloud applications~\cite{nastic2014provisioning}.
For business organisations,
logistics and asset management can be aided by the \ac{iot}~\cite{ding2013study}.
The \ac{iot} is characterised by its heterogeneity,
as an abundance of different devices can constitute an \ac{iot}.
Communication as well as embedded computation capabilities mark the common denominator on which \ac{iot} devices and networks depend.
While these features are a given in the age of constant mobile connectivity and open \acp{wlan},
they also constitute the vulnerabilities of \ac{iot} networks.
Being connected to networks leads to an increased attack surface.
Furthermore,
\ac{iot} devices are often cheap and manufactured in large numbers for short periods of time,
until they become obsolete.
After that,
newer versions are produced,
often for small prices.
Consequently,
re-use of hard- and software as well as low effort in programming contribute to insecure operating conditions~\cite{Spring.2016}.
The \ac{iot} is but a part of the development towards increased connectivity that inherently carries higher risks of cyber attacks.
Botnets targeting \ac{iot} devices,
such as \textit{Mirai}~\cite{kolias2017ddos},
industrial environments falling prey to attackers,
such as the power grid in the Ukraine in December 2015~\cite{Cherepanov.2017},
and ransomware attacks on healthcare infrastructure~\cite{slayton2018ransomware} and consumers~\cite{richardson2017ransomware} alike show the need for increased automated security in this brave new digital world.
As the networking paradigms are shifting from classic home and office networks to heterogeneous ad hoc networks,
security solutions have to adapt as well~\cite{Plaga.2019}.
Offensive security measures become an administrators friend to discover vulnerabilities along the attack phases.
By preemptive security,
such as vulnerability checking,
threats can be mitigated before an attacker can exploit them.
These capabilities are more relevant in the heterogeneous environments presented today.
Furthermore,
an insight regarding methods as well as tools of cyber criminals is becoming crucial for \ac{it} security professionals.
Since criminals often rely on publicly available tools,
an understanding of those allows security professionals to gain insight about the threat potential and possible attack vectors.
Furthermore,
if \ac{it} security professionals adapt to the methodology of a cyber criminal,
they obtain a new understanding of attacking a system,
potentially allowing for a more suited defense against attacks.
Additionally,
any vulnerability found by methods and tools of attackers is a vulnerability that can be mitigated before a real attack occurs.
The contribution of this work is
\begin{itemize}
    \item the identification and collection of the most well-known and used open source security assessment tools and
    \item the mapping of these tools to well-established attack models and.
    \item the analysis, comparison and discussion of the capabilities of these tools.
\end{itemize}
The remainder of this work is structured as follows.
The state of the art is presented in Section~\ref{sec:sota}. 
The methodology underlying this paper is introduced in Section~\ref{sec:methodology}.
The tools are introduced and evaluated in Section~\ref{sec:offsec_tools}.
A conclusion is drawn in Section~\ref{sec:conc}.

\section{State of the Art}
\label{sec:sota}
There is an abundance in literature regarding tools for offensive security purposes,
in numerous blogs,
but also in specialist books.
However,
an objective indication why the tools were chosen to be presented is not provided.
Either they are used and recommended by the author,
who usually is a security professional that has built a tool-box for themselves.
Or the tools are contained in a suite,
such as Kali Linux~\cite{kalilinux}.
\textit{Velu} presents the usage of Kali Linux for penetration testing in his book,
discussing the tools he deems most relevant~\cite{Velu.2016}.
\textit{Oakley} introduces red teaming in his book,
where tools are introduced in the respective stages based on experience of the author~\cite{Oakley.2019}.
\textit{Kim} presents practical penetration testing with the tools chosen in a similar fashion~\cite{Kim.2018}.
\textit{Forshaw} reduces the focus to tools for attacking network protocols,
thus setting a scope~\cite{Forshaw.2017}.
However,
the tools chosen are derived from his long experience.
In general,
it is beneficial to have tools introduced by professionals with a long experience in the fields,
as they took a long time to chose the right tool-box and become acquainted with it.
In this work,
however,
the focus is on creating tangible,
objective criteria for rating offensive security tools.

\section{Methodology}
\label{sec:methodology}
This section presents the methodology on which this work is founded.
First,
definitions of the terms are presented,
after that,
the sources from which the tools are collected.
Furthermore,
the scope of tools and applications is discussed,
a metric for attack stages is presented as well as the feature criteria of the tools.

\subsection{Definitons}
\label{ssec:definitions}
This paragraph presents the definitions of terms used in this work that are underlying to the evaluation. \par
\textit{Offensive information security}: Often called red teaming or penetration testing.
It is a concept that describes using tools and methods of an attacker to detect security vulnerabilities which then can be fixed before an attacker can exploit them. \par
\textit{Tool}: Finding a definition of the term tool in the context of software is exceedingly difficult.
For this work, 
a tool is defined as a software program that can be used as such without further software, 
except for operating system and corresponding environment.
A tool can consist of related parts that could be used in a stand alone-fashion,
but are distributed and commonly used together as they follow purposes along the path of a security assessment. \par
\textit{Freeware}: In the context of this work,
this term describes tools which can be obtained by private persons and professionals alike free of charge.
The free usage is not limited regarding the time of usage,
so trial versions of commercial software are not considered.
They are not specific to organisations. \par
\textit{Enterprise networks}: Consisting of \ac{it} infrastructures,
such as computers and servers.
Specifically excluded are \ac{ot} environments as found in industrial environments.

\subsection{Data Sources}
As the collection of exhaustive,
consistent lists of security tools with their attribution to a specific attack phase is difficult,
several sources were considered when identifying tools to evaluate in this work.
First,
the literature presented in Section~\ref{sec:sota} was used to extract the tools the authors used.
Second,
comprehensive lists that can be found online were employed.
The nmap project~\cite{nmap} provides a list of security tools,
called sectools~\cite{sectools}.
Furthermore,
\textit{r0lan} provides an overview of tools that is attributed to the phases they are used in~\cite{awesome-red-teaming}.
From these sources,
the most relevant tools were extracted.
Third,
well-known security distributions such as Kali~\cite{kalilinux} and Parrot~\cite{parrotlinux} Linux contain the tools that are most established in the security community.

\subsection{Scope}
The scope of this work are enterprise networks as discussed in Section~\ref{ssec:definitions},
consisting of computers, 
servers and auxiliary devices.
Furthermore,
the scope regarding the tools is limited on tools with a security focus.
Since there is a trend in security research as well as cyber criminals to use tools that are already installed on the target machine for exploitation purposes,
many tasks in security assessment can be performed without security-specific tools.
This technique is called living off the land.
An example is the use of Microsoft PowerShell for enumerating users and directories.

\subsection{Attack Metrics: MITRE ATT\&CK}
\label{sec:attack_metrics}
\begin{figure*}
    \includegraphics[width=\textwidth,height=7cm]{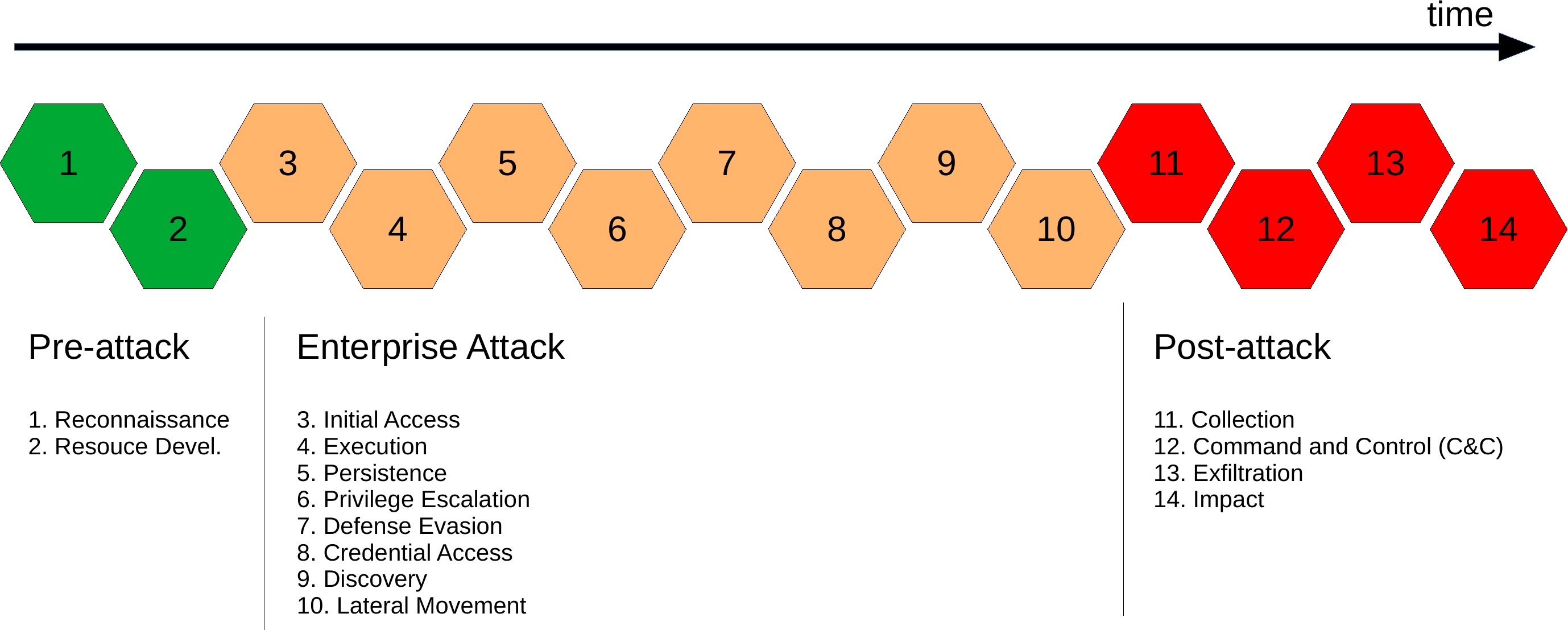}
  \caption{MITRE ATT\&CK Model}
  \label{fig:mitre}
\end{figure*}

The MITRE ATT\&CK matrix~\cite{mitre} was developed based on the well-established Lockheed Martin Cyber Kill Chain~\cite{cyberkillchain}.
Both aim at splitting a cyber attack into distinct phases during which an attacker follows a certain goal.
This is used to aid in comprehending the objectives of an attack and ultimately mitigating it.
The structure of the MITRE ATT\&CK model is shown in Figure~\ref{fig:mitre}.
Each of these phases requires a different set of tools,
so the aim of this work is mapping tools to these phases as discussed in Section~\ref{sec:offsec_tools}.
The phases used in the MITRE ATT\&CK Enterprise matrix are as follows,
with the description according to the MITRE-homepage~\cite{mitre}. \par
\textit{Reconnaissance}: The phase during which an adversary collects information about the target.
    Generally, 
    reconnaissance techniques are categorised in active,
    i.e. with the adversary interacting with the target system in an unexpected way,
    and passive,
    i.e. the adversary not directly interacting with the target system.\par
\textit{Resource Development}: The adversary is obtaining resources that can aid in attacking the system,
    such as accounts,
    systems,
    and other capabilities.
    The resources might be used in later phases,
    such as \ac{cc}.\par
\textit{Initial Access}: The adversary attempts to gain an initial foothold on the target system.
    This is the first phase with direct adversarial action on the target.\par
\textit{Execution}: The adversary executes malicious code on the target system.
    This code execution usually follows an underlying goal.
    Often,
    on-board capabilities of the target system,
    such as compilers and interpreters,
    aid in the execution of malicious code.\par
\textit{Persistence}: The phase in which the adversary aims to secure the foothold.
    Persistence allows re-entry and access to the system after the adversary logged out or the system rebooted.\par
\textit{Privilege Escalation}: In this phase,
    the adversary aims at obtaining higher privileges.
    Often,
    certain users are restricted from performing security-critical tasks,
    and the first foothold was performed with such restricted accounts.
    Elevating privileges allows the adversary to perform a wider variety of actions.\par
\textit{Defense Evasion}: After gaining access and elevating the privileges,
    the adversary actively evades detection by \ac{ids}.
    Obfuscation of the tools as well as deactivation of security measures are performed in this phase.\par
\textit{Credential Access}: The adversary aims at stealing account credentials for further use and to aid in following phases.\par
\textit{Discovery}: In this phase, 
    the adversary is gaining information about the environment in which the target system is located.
    This includes machines and services,
    accounts and users.\par
\textit{Lateral Movement}: In this phase,
    the adversary is moving through the target environment and infecting new systems.
    Often,
    the foothold with which entry to the network was gained does not contain the desired target,
    so lateral movement is necessary to reach devices that are not directly reachable from the outside.\par
\textit{Collection}: In this phase,
    the adversary gathers the desired information from the target machine.\par
\textit{Command and Control (C\&C)}: In this phase,
    the adversary executes control over the targeted systems and communicates with them.\par
\textit{Exfiltration}: The phase in which the adversary attempts to steal and obtain data without the owner of the target system noticing.
    The data has to be sent in a fashion that does not cause suspicion.\par
\textit{Impact}: In this phase,
    the adversary maliciously impacts the target system by destroying or restricting its functionality.
    This activity can easily be detected by the owner of the system.\par

\subsection{Tool Features}
In order to evaluate and rate the tools,
a metric needs to be defined.
This metric should contain tangible,
verifiable features.
The features used in this work are listed as follows:
\begin{itemize}
    \item Actively maintained: This feature is evaluated according to the latest release and the average number of releases per year. Actively maintained tools provide bug fixes and the integration of new features and protocols as well as a more active support.
    \item Usage: This feature discusses the licence a tool is published under, the support a user can expect and whether or not a paid version of the tool is available.
    \item Technical: This feature describes the interface of the tool for a user as well as the programming language the tool is programmed in. This is important as the way of interaction can make a tool more difficult or easier to apply, while the programming language in open source tools describes whether or not a user could adapt and extend the tool.
\end{itemize}
These features allow an assessment of the tools according to several dimensions.
It can be derived if the tool is actively developed and likely to be adapted to new technologies.
Furthermore,
the capabilities for extending and embedding the tool into a toolchain can be obtained from these features.

\section{Analysis}
\label{sec:offsec_tools}
In this section the relevant tools are identified and the metric is applied for comparison.
Then, the results for the comparison are discussed.

\subsection{Identification}
For the evaluation,
well-established,
commonly used security tools tailored for each of the phases as presented in Section~\ref{sec:attack_metrics} were identified,
collected and compared.
As it is not trivial to find an exhaustive overview of security tools,
several sources were used to obtain information.
Apart from books presented in Section~\ref{sec:sota},
web resources were considered as well.
The nmap network scanner~\cite{nmap} hosts a list of security tools~\cite{sectools}.
Every tool that fits the scope of this work in the top 50 tools of the sectools-list is evaluated in this work.
Thus,
a good coverage of relevant tools is expected.
Furthermore,
\textit{r0lan} provides an overview of tools commonly used for the individual phases~\cite{awesome-red-teaming}.
Among links to tools,
\textit{r0lan} provides an abundance of sources that describe methods rather than tools,
and information how to employ tools without a first focus on security,
such as PowerShell,
to perform security-relevant tasks.
From these sources,
the most used tools were extracted.
They are listed in Table~\ref{tab:methodology_vulnerabilities_siemens_overview},
categorised according to the phases of the MITRE ATT\&CK metric.

\subsection{Comparison}
The tools identified in the previous subsection are listed in Table~\ref{tab:methodology_vulnerabilities_siemens_overview},
with the criteria alongside which they are evaluated.
\begin{table*}[ht]
		\renewcommand{\arraystretch}{1.3}
		\caption{Security Assessment Tools Categorised According to the Phases They Are Used During}
		\label{tab:methodology_vulnerabilities_siemens_overview}
		\centering
		\scriptsize
		\begin{tabular}{l c c c c c c c c c c c}
			\toprule
			\textbf{Tools} & \phantom{a} & \multicolumn{3}{c}{\textbf{Releases}} & \phantom{a} & \multicolumn{3}{c}{\textbf{Usage aspects}} & \phantom{a} & \multicolumn{2}{c}{\textbf{Technical aspects}} \\
			&             &         First & p.a. & Latest &             & Licence & Supp. & Paid & & Interf. & Lang.  \\
			\cmidrule{1-1} \cmidrule{3-5} \cmidrule{7-9} \cmidrule{11-12} 
			\textbf{Reconnaissance} \\
			nmap~\cite{nmap} & & 1997 & 10 & 2020 & & Nmap Publ. Src. & Forum & no & & GUI, CLI & Lua, C, C++, Python, Shell  \\
			OpenVAS~\cite{openvas} & & 2006 & 9 & 2020 & & GNU GPLv2 2.0 & Forum & no & & GUI, CLI & C, NASL, Yacc, Shell, C++  \\
			Maltego~\cite{maltego} & & 2007 & 11 & 2020 & & Proprietary & Forum & yes & & GUI & Java  \\
			Shodan~\cite{shodan} & & 2009 & n/a & 2020 & & Proprietary & Forum & yes & & WUI, CLI & n/a   \\
			TheHarvester~\cite{theharvester} & & 2011 & 3 & 2020 & & GPLv2 & Forum & no & & CLI & Python   \\
			\cmidrule{1-1} \cmidrule{3-5} \cmidrule{7-9} \cmidrule{11-12}
			\textbf{Initial Access} \\
			Aircrack-ng~\cite{aircrackng} & & 2006 & 2 & 2020 & & GPLv2 & Forum & no & & GUI, CLI & C, M4, C\#, Shell, Python, Roff   \\
			GoPhish~\cite{gophish} & & 2013 & 2 & 2020 & & MIT & GitHub & no & & WUI, CLI & Go, JavaScript   \\
			msf~\cite{msf} & & 2014 & 2 & 2020 & & Apache 2.0 & Forum & no & & GUI, CLI & C++ \\
			\cmidrule{1-1} \cmidrule{3-5} \cmidrule{7-9} \cmidrule{11-12}
			\textbf{Persistence} \\
			C99 webshell~\cite{c99webshell} & & 2005 & n/a & n/a & & n/a & n/a & no & & CLI & PHP   \\
			Reptile~\cite{reptile} & & 2018 & 1 & 2020 & & n/a & n/a & no & & CLI & C, C++, Yacc, Perl, Shell   \\
			SharPersist~\cite{sharpersist} & & 2019 & 1 & 2020 & & Apache 2.0 & Forum & no & & CLI & C\#  \\
			\cmidrule{1-1} \cmidrule{3-5} \cmidrule{7-9} \cmidrule{11-12}
			\textbf{Privilege Escalation} \\
			searchsploit~\cite{searchsploit} & & 2014 & n/a & 2020 & & GPL-2.0 & Forum & no & & CLI & C, Python, Ruby, Perl, PHP \\
			msf~\cite{msf} & & 2014 & 2 & 2020 & & Apache 2.0 & Forum & yes & & GUI, CLI & C++   \\
			\cmidrule{1-1} \cmidrule{3-5} \cmidrule{7-9} \cmidrule{11-12}
			\textbf{Defense Evasion} \\
			veil-evasion~\cite{veil-evasion} & & 2015 & 21 & 2016 & & GPL-3.0 & Forum & no & & CLI & Python, C, Shell, C++ \\
			shellter~\cite{shellter} & & 2014 & 14.7 & 2017 & & special & Forum & yes & & CLI & n/a  \\
			msf~\cite{msf} & & 2014 & 2 & 2020 & & Apache 2.0 & Forum & yes & & GUI, CLI & C++   \\
			\cmidrule{1-1} \cmidrule{3-5} \cmidrule{7-9} \cmidrule{11-12}
			\textbf{Credential Access} \\
			LaZange~\cite{lazagne} & & 2015 & 3.6 & 2019 & & LGPL-3.0 & Forum & no & & CLI & Python \\
			Responder~\cite{responder} & & 2014 & 7.5 & 2015\footnotemark & & GPL-3.0 & Forum & no & & CLI & Python \\
			Mimikatz~\cite{mimikatz} & & 2007 & 0.5 & 2020 & & CC-BY 4.0 & Forum & no & & CLI & C \\
			msf~\cite{msf} & & 2014 & 2 & 2020 & & Apache 2.0 & Forum & no & & GUI, CLI & C++ \\
			\cmidrule{1-1} \cmidrule{3-5} \cmidrule{7-9} \cmidrule{11-12}
			\textbf{Discovery} \\
			linuxprivchecker~\cite{linuxprivchecker} & & 2015 & n/a & 2020 & & - & Forum & no & & CLI & Python \\
			windows-privesc-check~\cite{windows-privesc-check} & & 2010 & n/a & 2015 & & GPL-2.0 & Forum & no & & CLI & Python \\
			Bloodhound~\cite{bloodhound} & & 2016 & 7 & 2020 & & GPL-3.0 & Forum & no & & GUI, CLI & C++ \\
			\cmidrule{1-1} \cmidrule{3-5} \cmidrule{7-9} \cmidrule{11-12}
			\textbf{Lateral Movement} \\
			Mimikatz~\cite{mimikatz} & & 2007 & 0.5 & 2020 & & CC-BY 4.0 & Forum & no & & CLI & C \\
			Rubeus~\cite{rubeus} & & 2018 & n/a & 2020 & & BSD 3-clause & Forum & no & & CLI & C\# \\
			msf~\cite{msf} & & 2014 & 2 & 2020 & & Apache 2.0 & Forum & yes & & GUI, CLI & C++ \\
			\cmidrule{1-1} \cmidrule{3-5} \cmidrule{7-9} \cmidrule{11-12}
			\textbf{Collection} \\
			Veil-Pillage~\cite{veil-pillage} & & 2014 & n/a & 2015 & & GPL-3.0 & Forum & no & & CLI & PowerShell, Python \\
			msf~\cite{msf} & & 2014 & 2 & 2020 & & Apache 2.0 & Forum & yes & & GUI, CLI & C++ \\
			\cmidrule{1-1} \cmidrule{3-5} \cmidrule{7-9} \cmidrule{11-12}
			\textbf{Command and Control} \\
			empire~\cite{empire} & & 2015 & 4.33 & 2018\footnotemark[\value{footnote}] & & BSD 3-clause & Forum & no & & CLI & PowerShell, Python \\
			msf~\cite{msf} & & 2014 & 2 & 2020 & & Apache 2.0 & Forum & yes & & GUI, CLI & C++ \\
			\cmidrule{1-1} \cmidrule{3-5} \cmidrule{7-9} \cmidrule{11-12}
			\textbf{Exfiltration} \\
			DET~\cite{det} & & 2016 & n/a & 2019 & & MIT & Forum & no & & CLI & Python \\
			Cloakify-Factory~\cite{cloakifyfactory} & & 2018 & 4 & 2018 & & MIT & Forum & no & & GUI, CLI & Python \\
			msf~\cite{msf} & & 2014 & 2 & 2020 & & Apache 2.0 & Forum & yes & & GUI, CLI & C++ \\
			\cmidrule{1-1} \cmidrule{3-5} \cmidrule{7-9} \cmidrule{11-12}
			\textbf{Impact} \\
			Veil-Pillage~\cite{veil-pillage} & & 2014 & n/a & 2015 & & GPL-3.0 & Forum & no & & CLI & PowerShell, Python \\
			msf~\cite{msf} & & 2014 & 2 & 2020 & & Apache 2.0 & Forum & yes & & GUI, CLI & C++ \\
			\bottomrule
		\end{tabular}
	\end{table*}
	\footnotetext{These tools are indicated to be no longer actively maintained.}
It can be seen that the first stage has the most extensive number of tools available.
Most tools are still actively maintained,
meaning new releases are provided at the time of this work.
However,
a few tools,
empire~\cite{empire} and Responder~\cite{responder},
are indicated to be deprecated.
Furthermore,
several tools, 
Veil-Evasion~\cite{veil-evasion},
Veil-Pillage~\cite{veil-pillage}
shellter~\cite{shellter},
windows-privesc-check~\cite{windows-privesc-check},
and Cloakify-Facory~\cite{cloakifyfactory} have their latest releases older than a year,
which indicates limited maintenance of these tools.
Since most versions are free,
do not have a paid version and are developed by members of the community,
most support is provided in terms of forums or mailing lists.
Every tool provides a \ac{cli},
some tools additionally provide a \ac{gui} or \ac{wui}.
The \ac{cli}-capabilities allow for the tool to be integrated into toolchains,
with the output being piped into other applications.
Furthermore,
since a number of tools is written in Python,
integration of the source code into user tools is easily possible.
All tools,
except for Maltego~\cite{maltego} and shellter~\cite{shellter},
provide their code for a user to extend and adapt.
	
\subsection{Discussion}
The features based on which the tools are evaluated are intended to provide information for a user to pick a tool that fits her need.
Furthermore,
the quality and suitability of tools,
as well as their versatility is evaluated.
For example,
msf~\cite{msf} can be used in nine of 14 phases,
making it the most versatile and powerful tool in this comparision.
This is due to the toolbox-approach that allows msf to load different modules for specific tasks.
Furthermore,
research underlying this work shows that some phases have significantly more tools created for them than others.
For example,
the execution phase does not have a singular tool dedicated to it.
Instead,
board measures of the target system are used,
or tools that are not security-specific,
such as web servers or communication tools.
Furthermore,
the variety of tools,
especially in the reconnaissance-phase,
is an indicator of knowledge and experience an \ac{it} security professional should have.

\section{Conclusion}
\label{sec:conc}
This work highlights a few insights.
First,
there is a plethora of \ac{it} security assessment tools available that can be used by professionals as well as cyber criminals.
Being aware of these tools and gaining familiarity is therefore crucial for security experts.
The majority is freely available and actively maintained,
with examples and help readily available.
Second,
some phases have more tools dedicated to them than others.
Reconnaissance has an abundance of publicly available tools solely for the purpose of gathering information that can be used to exploit a target.
The initial access phase has several tools as well,
as attacks can be aimed at different types of targets.
For example,
there is a number of tools to attack websites,
a different set of tools for attacking databases and further tools for other attack vectors.
Understanding these as a professional is crucial for hardening any potential attack vector and preventing attacks from happening in the first place.
Other phases rely on tools already available on the target systems,
such as programming compilers and interpreters,
PowerShell,
netcat and other tools.
These can be misused to perform malicious activity,
a fact of which \ac{it} security professionals need to be aware of as well.
This living off the land can be discussed in a future work,
as there is an abundance of tools that can be misused under the right circumstances.
Furthermore,
tools for cracking passwords and for monitoring and exploiting wireless devices were not considered in this work,
as they would exceed the scope.
An exhaustive overview of such tools can be discussed in a future work as well.
This work shows that gaining insight about attacks is crucial for defense.
A method commonly used to obtain information about attackers,
their tools and aims is honeypots~\cite{Fraunholz.2017c,Fraunholz.2017h}.
Attributing attackers and attacks is similarly important in order to implement counter measures suitable for the attacks which are identified as most likely or having most impact~\cite{Fraunholz.2017f}.
In general,
detecting attacks is becoming increasingly difficult due to the changes in \ac{it} infrastructure.
Using context information~\cite{Duque_Anton.2017c},
employing novel machine learning approaches~\cite{anton2019anomaly} and creating and providing data to train \acp{ids}~\cite{Duque_Anton.2019a} on are required to secure \ac{it} network against current and future attacks.

\bibliographystyle{IEEEtran}
\bibliography{literature}

\begin{thebibliography}{10}
\providecommand{\url}[1]{#1}
\csname url@samestyle\endcsname
\providecommand{\newblock}{\relax}
\providecommand{\bibinfo}[2]{#2}
\providecommand{\BIBentrySTDinterwordspacing}{\spaceskip=0pt\relax}
\providecommand{\BIBentryALTinterwordstretchfactor}{4}
\providecommand{\BIBentryALTinterwordspacing}{\spaceskip=\fontdimen2\font plus
\BIBentryALTinterwordstretchfactor\fontdimen3\font minus
  \fontdimen4\font\relax}
\providecommand{\BIBforeignlanguage}[2]{{%
\expandafter\ifx\csname l@#1\endcsname\relax
\typeout{** WARNING: IEEEtran.bst: No hyphenation pattern has been}%
\typeout{** loaded for the language `#1'. Using the pattern for}%
\typeout{** the default language instead.}%
\else
\language=\csname l@#1\endcsname
\fi
#2}}
\providecommand{\BIBdecl}{\relax}
\BIBdecl

\bibitem{formisano2015advantages}
C.~Formisano, D.~Pavia, L.~Gurgen, T.~Yonezawa, J.~A. Galache, K.~Doguchi, and
  I.~Matranga, ``The advantages of iot and cloud applied to smart cities,'' in
  \emph{2015 3rd International Conference on Future Internet of Things and
  Cloud}.\hskip 1em plus 0.5em minus 0.4em\relax IEEE, 2015, pp. 325--332.

\bibitem{nastic2014provisioning}
S.~Nastic, S.~Sehic, D.-H. Le, H.-L. Truong, and S.~Dustdar, ``Provisioning
  software-defined iot cloud systems,'' in \emph{2014 international conference
  on future internet of things and cloud}.\hskip 1em plus 0.5em minus
  0.4em\relax IEEE, 2014, pp. 288--295.

\bibitem{ding2013study}
W.~Ding, ``Study of smart warehouse management system based on the iot,'' in
  \emph{Intelligence computation and evolutionary computation}.\hskip 1em plus
  0.5em minus 0.4em\relax Springer, 2013, pp. 203--207.

\bibitem{Spring.2016}
\BIBentryALTinterwordspacing
T.~Spring, ``Code reuse a peril for secure software development,''
  \emph{threatpost}, 2016. [Online]. Available:
  \url{https://threatpost.com/code-reuse-a-peril-for-secure-software-development/122476/}
\BIBentrySTDinterwordspacing

\bibitem{kolias2017ddos}
C.~Kolias, G.~Kambourakis, A.~Stavrou, and J.~Voas, ``Ddos in the iot: Mirai
  and other botnets,'' \emph{Computer}, vol.~50, no.~7, pp. 80--84, 2017.

\bibitem{Cherepanov.2017}
A.~Cherepanov, ``{Win32/Industroyer} - a new threat for industrial control
  systems,'' ESET, Tech. Rep., June 2017.

\bibitem{slayton2018ransomware}
T.~B. Slayton, ``Ransomware: the virus attacking the healthcare industry,''
  \emph{Journal of Legal Medicine}, vol.~38, no.~2, pp. 287--311, 2018.

\bibitem{richardson2017ransomware}
R.~Richardson and M.~M. North, ``Ransomware: Evolution, mitigation and
  prevention,'' \emph{International Management Review}, vol.~13, no.~1, p.~10,
  2017.

\bibitem{Plaga.2019}
S.~Plaga, N.~Wiedermann, S.~Duque~Anton, S.~Tatschner, H.~D. Schotten, and
  T.~Newe, ``Securing future decentralised industrial {IoT} infrastructures:
  Challenges and free open source solutions,'' \emph{Future Generation Computer
  Systems}, vol.~93, pp. 596--608, 2019.

\bibitem{kalilinux}
\BIBentryALTinterwordspacing
{Offensive Security}, ``Our most advanced penetration testing distribution,
  ever.'' last visited 03-11-2020. [Online]. Available:
  \url{https://www.kali.org/}
\BIBentrySTDinterwordspacing

\bibitem{Velu.2016}
V.~K. Velu, \emph{Mastering Kali Linux for Advanced Penetration Testing},
  2nd~ed.\hskip 1em plus 0.5em minus 0.4em\relax Birmingham, Mumbay: packt,
  2016.

\bibitem{Oakley.2019}
J.~G. Oakyley, \emph{Professional Red Teaming}.\hskip 1em plus 0.5em minus
  0.4em\relax Apress, 2019.

\bibitem{Kim.2018}
P.~Kim, \emph{The Hacker Playbook 3}.\hskip 1em plus 0.5em minus 0.4em\relax
  Secure Planet, 2018.

\bibitem{Forshaw.2017}
J.~Forshaw, \emph{Attacking Network Protocols}.\hskip 1em plus 0.5em minus
  0.4em\relax San Francisco: no starch press, 2017.

\bibitem{nmap}
\BIBentryALTinterwordspacing
G.~Lyon, ``nmap: the network mapper,'' last visited 06-11-2020. [Online].
  Available: \url{https://nmap.org/}
\BIBentrySTDinterwordspacing

\bibitem{sectools}
\BIBentryALTinterwordspacing
------, ``{Top Network Security Tools},'' last visited 06-11-2020. [Online].
  Available: \url{https://sectools.org/}
\BIBentrySTDinterwordspacing

\bibitem{awesome-red-teaming}
\BIBentryALTinterwordspacing
{r0lan}, ``{Awesome-Red-Teaming},'' last visited 05-11-2020. [Online].
  Available: \url{https://github.com/yeyintminthuhtut/Awesome-Red-Teaming}
\BIBentrySTDinterwordspacing

\bibitem{parrotlinux}
\BIBentryALTinterwordspacing
{Parrot Security CIC}, ``Parrot os,'' last visited 03-11-2020. [Online].
  Available: \url{https://parrotlinux.org/}
\BIBentrySTDinterwordspacing

\bibitem{mitre}
\BIBentryALTinterwordspacing
``Mitre att\&ck enterprise matrix,'' last visited 03-11-2020. [Online].
  Available: \url{https://attack.mitre.org/matrices/enterprise/}
\BIBentrySTDinterwordspacing

\bibitem{cyberkillchain}
\BIBentryALTinterwordspacing
{Lokheed Martin}, ``The cyber kill chain,'' last visited 03-11-2020. [Online].
  Available:
  \url{https://www.lockheedmartin.com/en-us/capabilities/cyber/cyber-kill-chain.html}
\BIBentrySTDinterwordspacing

\bibitem{openvas}
\BIBentryALTinterwordspacing
{Greenbone}, ``{OpenVAS - Open Vulnerability Assessment Scanner},'' last
  visited 06-11-2020. [Online]. Available:
  \url{https://www.openvas.org/index-de.html}
\BIBentrySTDinterwordspacing

\bibitem{maltego}
\BIBentryALTinterwordspacing
{Paterva}, ``{Maltego},'' last visited 06-11-2020. [Online]. Available:
  \url{https://www.maltego.com/}
\BIBentrySTDinterwordspacing

\bibitem{shodan}
\BIBentryALTinterwordspacing
{Shodan}, ``{The search engine for the web},'' last visited 06-11-2020.
  [Online]. Available: \url{https://www.shodan.io/}
\BIBentrySTDinterwordspacing

\bibitem{theharvester}
\BIBentryALTinterwordspacing
C.~Martorella, ``{theHarvester},'' last visited 06-11-2020. [Online].
  Available: \url{https://github.com/laramies/theHarvester}
\BIBentrySTDinterwordspacing

\bibitem{aircrackng}
\BIBentryALTinterwordspacing
{Aircrack-ng}, ``Aircrack-ng,'' last visited 06-11-2020. [Online]. Available:
  \url{https://www.aircrack-ng.org/}
\BIBentrySTDinterwordspacing

\bibitem{gophish}
\BIBentryALTinterwordspacing
J.~Wright, ``Open-source phishing framework,'' last visited 06-11-2020.
  [Online]. Available: \url{https://getgophish.com/}
\BIBentrySTDinterwordspacing

\bibitem{msf}
\BIBentryALTinterwordspacing
{Rapid7}, ``{metasploit - The world’s most used penetration testing
  framework},'' last visited 06-11-2020. [Online]. Available:
  \url{https://www.metasploit.com/}
\BIBentrySTDinterwordspacing

\bibitem{c99webshell}
\BIBentryALTinterwordspacing
M.~Cermak, ``C99 webshell with php7 and mysql support,'' last visited
  06-11-2020. [Online]. Available:
  \url{https://github.com/cermmik/C99-WebShell}
\BIBentrySTDinterwordspacing

\bibitem{reptile}
\BIBentryALTinterwordspacing
I.~Augusto, ``Reptile,'' last visited 06-11-2020. [Online]. Available:
  \url{https://github.com/f0rb1dd3n/Reptile}
\BIBentrySTDinterwordspacing

\bibitem{sharpersist}
\BIBentryALTinterwordspacing
B.~Hawkins, ``Sharpersist,'' last visited 06-11-2020. [Online]. Available:
  \url{https://github.com/fireeye/SharPersist}
\BIBentrySTDinterwordspacing

\bibitem{searchsploit}
\BIBentryALTinterwordspacing
{Offensive Security}, ``{The Exploit Database Git Repository},'' last visited
  05-11-2020. [Online]. Available:
  \url{https://github.com/offensive-security/exploitdb}
\BIBentrySTDinterwordspacing

\bibitem{veil-evasion}
\BIBentryALTinterwordspacing
{The Veil-Framework}, ``{Veil-Evasion},'' last visited 05-11-2020. [Online].
  Available: \url{https://github.com/Veil-Framework/Veil-Evasion}
\BIBentrySTDinterwordspacing

\bibitem{shellter}
\BIBentryALTinterwordspacing
{Economou, Kyriakos}, ``{Shellter},'' last visited 05-11-2020. [Online].
  Available: \url{https://www.shellterproject.com/}
\BIBentrySTDinterwordspacing

\bibitem{lazagne}
\BIBentryALTinterwordspacing
{AlessandroZ}, ``{LaZagne},'' last visited 05-11-2020. [Online]. Available:
  \url{https://github.com/AlessandroZ/LaZagne}
\BIBentrySTDinterwordspacing

\bibitem{responder}
\BIBentryALTinterwordspacing
{SpiderLabs}, ``{Responder},'' last visited 05-11-2020. [Online]. Available:
  \url{https://github.com/SpiderLabs/Responder}
\BIBentrySTDinterwordspacing

\bibitem{mimikatz}
\BIBentryALTinterwordspacing
B.~Delpy, ``{mimikatz},'' last visited 05-11-2020. [Online]. Available:
  \url{https://github.com/gentilkiwi/mimikatz}
\BIBentrySTDinterwordspacing

\bibitem{linuxprivchecker}
\BIBentryALTinterwordspacing
M.~Contino, ``{linuxprivchecker},'' last visited 05-11-2020. [Online].
  Available: \url{https://github.com/sleventyeleven/linuxprivcheckers}
\BIBentrySTDinterwordspacing

\bibitem{windows-privesc-check}
\BIBentryALTinterwordspacing
{pentestmonkey}, ``{windows-privesc-check},'' last visited 05-11-2020.
  [Online]. Available:
  \url{https://github.com/pentestmonkey/windows-privesc-check}
\BIBentrySTDinterwordspacing

\bibitem{bloodhound}
\BIBentryALTinterwordspacing
{BloodHoundAD}, ``{BloodHound},'' last visited 05-11-2020. [Online]. Available:
  \url{https://github.com/BloodHoundAD/BloodHound}
\BIBentrySTDinterwordspacing

\bibitem{rubeus}
\BIBentryALTinterwordspacing
{GhostPack}, ``{Rubeus},'' last visited 05-11-2020. [Online]. Available:
  \url{https://github.com/GhostPack/Rubeus}
\BIBentrySTDinterwordspacing

\bibitem{veil-pillage}
\BIBentryALTinterwordspacing
{The Veil-Framework}, ``{Veil-Pillage},'' last visited 05-11-2020. [Online].
  Available: \url{https://github.com/Veil-Framework/Veil-Pillage}
\BIBentrySTDinterwordspacing

\bibitem{empire}
\BIBentryALTinterwordspacing
{EmpireProject}, ``{Empire},'' last visited 05-11-2020. [Online]. Available:
  \url{https://github.com/EmpireProject/Empire}
\BIBentrySTDinterwordspacing

\bibitem{det}
\BIBentryALTinterwordspacing
{PaulSec}, ``{DET (extensible) Data Exfiltration Toolkit},'' last visited
  05-11-2020. [Online]. Available: \url{https://github.com/PaulSec/DET}
\BIBentrySTDinterwordspacing

\bibitem{cloakifyfactory}
\BIBentryALTinterwordspacing
{TryCatchHCF}, ``{Cloakify},'' last visited 05-11-2020. [Online]. Available:
  \url{https://github.com/TryCatchHCF/Cloakify}
\BIBentrySTDinterwordspacing

\bibitem{Fraunholz.2017c}
D.~Fraunholz, M.~Zimmermann, A.~Hafner, and H.~D. Schotten, ``Data mining in
  long-term honeypot data,'' \emph{IEEE International Conference on Data Mining
  series - Workshop on Data Mining for Cyber-Security}, 2017.

\bibitem{Fraunholz.2017h}
D.~Fraunholz, M.~Zimmermann, S.~{Duque Anton}, J.~Schneider, and H.~D.
  Schotten, ``Distributed and highly-scalable wan network attack sensing and
  sophisticated analysing framework based on honeypot technology,''
  \emph{International Conference on Cloud Computing, Data Science {\&}
  Engineering}, vol.~7, 2017.

\bibitem{Fraunholz.2017f}
D.~Fraunholz, D.~Krohmer, S.~{Duque Anton}, and H.~D. Schotten, ``Yaas - on the
  attribution of honeypot data,'' \emph{International Journal on Cyber
  Situational Awareness}, vol.~2, no.~1, pp. 31--48, 2017.

\bibitem{Duque_Anton.2017c}
S.~Duque~Anton, D.~Fraunholz, S.~Teuber, and H.~D. Schotten, ``A question of
  context: Enhancing intrusion detection by providing context information,'' in
  \emph{13th Conference of Telecommunication, Media and Internet
  Techno-Economics (CTTE-17)}, 2017.

\bibitem{anton2019anomaly}
S.~D. Duque~Anton, S.~Sinha, and H.~D. Schotten, ``Anomaly-based intrusion
  detection in industrial data with {SVM} and random forests,'' in \emph{2019
  International Conference on Software, Telecommunications and Computer
  Networks (SoftCOM)}.\hskip 1em plus 0.5em minus 0.4em\relax IEEE, 2019, pp.
  1--6.

\bibitem{Duque_Anton.2019a}
S.~Duque~Anton, M.~Gundall, D.~Fraunholz, and H.~D. Schotten, ``Implementing
  scada scenarios and introducing attacks to obtain training data for intrusion
  detection methods,'' in \emph{International Conference on Cyber Warfare and
  Security (ICCWS)}, 2019.

\end{thebibliography}

\end{document}